\documentclass[twocolumn,aps,prx,10pt,superscriptaddress]{revtex4-2}

\usepackage{graphicx}
\usepackage{amsmath,enumitem}
\usepackage{mathtools}
\usepackage{amssymb}
\usepackage[dvipsnames]{xcolor}
\usepackage{seqsplit}

\usepackage{url}

\usepackage{breakurl}
\usepackage[breaklinks]{hyperref}

\begin{document}

\title{Some recent advances in urban system science: models and data}

\author{Elsa Arcaute}\email{e.arcaute@ucl.ac.uk} \affiliation{Centre for Advanced Spatial Analysis, University College London, 90 Tottenham Court Road, London, W1T 4TJ, UK}

\author{Jos\'e J. Ramasco}\email{jramasco@ifisc.uib-csic.es} \affiliation{Instituto de F\'{\i}sica Interdisciplinar y Sistemas Complejos IFISC (CSIC-UIB), Campus UIB, 07122 Palma de Mallorca, Spain}

\begin{abstract}
Cities are characterized by the presence of a dense population with a high potential for interactions between individuals of diverse backgrounds. They appear in parallel to the Neolithic revolution a few millennia ago. The advantages brought in terms of agglomeration for economy, innovation, social and cultural advancements have kept them as a major landmark in recent human history. There are many different aspects to study in urban systems from a scientific point of view, just to name a few one can concentrate in demography and population evolution, mobility, economic output, land use and urban planning, home accessibility and real estate market, energy and water consumption, waste processing, health, education, integration of minorities, etc. In the last decade, the introduction of communication and information technologies have enormously facilitated the collection of datasets on these and other questions, making possible a more quantitative approach to city science. All these topics have been addressed in many works in the literature, and we do not intend to offer here a systematic review. 
Instead, we will only provide a brief taste of some of these above-mentioned aspects, which could serve as an introduction to a subsequent special number. 
Such a non-systematic view will lead us to leave outside many relevant papers, and for this we apologise. 
\end{abstract}

\maketitle

\section*{Introduction}

Cities, and urban systems in general, present generic patterns despite being the result of a diverse set of processes and constraints. In this section, we look at some of the old and more recent attempts to encapsulate these within a mathematical framework. For example, agglomeration effects have been observed worldwide for over a century. This has led researchers to investigate further the effects of city size on urban indicators. The topic of urban scaling laws sprouting from this initial idea, has generated a lot of activity, and through the scrutinization of the limitations of the method, new paradigms and models have emerged. The framework on the other hand, does not consider the heterogeneities within cities, nor the mechanisms giving rise to the observed urban metrics. And although many of these are still an open problem, these processes take place within the spatial fabric of the city, and their physical embeddedness cannot be disentangled from their effect. In this sense, many of the spatial correlations of the different processes taking place in cities, are tightly related to the spatial distribution of functions and transport, which are both closely linked to the morphology of cities. Such an interdependency is yet to be understood. Advancing this field necessitates a quantification of the \emph{form} of the city, and although all cities look different, they all reveal fractal patterns. 
These patterns play a role in modulating the intensity of the interactions between functions. Sometimes, the modulating distance is not necessarily physical, and corresponds to a proxy of higher probability of interaction due to similarity or complementarity between the components.
And it is the different intensities of interactions within systems and across scales that gives rise to another generic pattern observed in most complex systems: a hierarchical organisation.

\section*{Urban scaling laws: what is missing?}

Cities can be thought of structures created as a convergent solution to sustain the many necessities of human beings. By agglomerating in an area, individuals have been able to share resources and facilitate exchanges allowing for better productivity. Over time, different processes, such as trade, skill matching and specialisation, and the evolution of transport to mention a few, have come together shaping the spatial distributions of land uses in cities. 
Different models attempting to explain the observed patterns have been proposed since the 19th century, such as von Th\"unen’s model of concentric rings of land uses as detailed in his treatise of \emph{The Isolated State} in 1826, Christaller’s Central Place Theory \cite{christaller_central_1933} aiming at explaining a hierarchical order in the distribution of settlement sizes and their functions, and L{\"o}sch’s location theory \cite{losch_economics_1954} where he emphasized that transport cannot be disentangled to the observed agglomerations. Overall, for more than half a century, there have been many proposals looking at cities from the perspective of flows \cite{Haggett_Chorley1969}, and of complexity science through Berry’s proposal to consider cities as “systems of systems” \cite{berry_cities_1964}. It is beyond the scope of this paper to provide a comprehensive review of the different theoretical and modelling frameworks to cities, it is nevertheless fundamental to mention two of the pioneers and driving forces behind the development of a science of cities within the framework of complexity science: Denise Pumain \cite{pumain_pour_1997,pumain_les_1992} and Michael Batty \cite{batty_new_2013,batty_inventing_2018}.

Although the quest to model cities started more than a century ago, capturing all the processes and their interdependencies continues to be a challenge. On the other hand, emergent patterns, such as agglomeration economies, have been identified since a century ago \cite{marshall_a._principles_1890}, and continue to be investigated \cite{duranton_2004_micro,glaeser_introduction_2010}. These ideas were extended beyond the realm of economics, and simplified through the following relationship: $O \sim P^{\beta}$, where the output $O$ corresponds to an urban indicator, $P$ to the population of a city, and the exponent $\beta$ indicates if there are effects due to concentration of people, i.e. if $\beta > 1$ the output $O$ is more than proportional to the amount of people $P$ in the city \cite{bettencourt_growth_2007}. A cautionary note on the observed output was brought forward by Denise Pumain \cite{pumain_scaling_2004,pumain_evolutionary_2006}. Within the industrial sector, the value of the exponent will not always be larger than $1$, but it will depend on the level of maturity of the sector. For example, at the stage when an industry is producing an innovation, the activity is mostly concentrated in big cities, and therefore the expected agglomeration effects with $\beta > 1$ will be observed. After this initial adoption period, the exponent will shift to $\beta \sim 1$, indicating that productivity is no longer concentrated in big cities, but has diffused to smaller ones where production might be cheaper due to lower rents and wages. Similar findings related to the value of the exponent with the phase of economic growth can be found in \cite{strano_rich_2016}.

Scaling laws in urban systems attracted a lot of interest in many different areas \cite{samaniego_cities_2008, gomez-lievano_statistics_2012,ribeiro_model_2017,meirelles_evolution_2018,cabrera-arnau_effect_2020}. Nevertheless, some contradictions started to emerge, touching upon important questions related to the role of city size: are larger cities greener \cite{oliveira_large_2014,fragkias_does_2013,fuller_scaling_2009}? Are they more congested \cite{louf_how_2014,depersin_global_2018}? And are larger cities more unequal \cite{sarkar_urban_2018}? In addition, limitations of the method were also identified: does the result depend on the definition of the system of cities \cite{arcaute_constructing_2015}? Or are these contradictions the result of a poor statistical modelling framework \cite{leitao_is_2016,shalizi_scaling_2011,gerlach_testing_2019,corral_truncated_2020}? Furthermore, the reconciliation between transversal and longitudinal scaling to better understand the evolution of cities is yet to be discerned \cite{hong_common_2018P,ribeiro_relation_2020}. It is also expected, that such laws would hold for past and contemporary settlements \cite{ortman_cities_2020}. Overall, one of the most problematic aspects of the proposed urban scaling laws, is that they do not consider the interaction between cities \cite{altmann_spatial_2020}, nor their correlations \cite{ribeiro_association_2021}. 
It is well-known within the study of out of equilibrium systems that power laws might arise from single processes leading to a homogeneous relationship such as the simplistic equation described above \cite{meakin_fractals_1998}. Nevertheless, when many correlated processes intervene, homogeneity is lost. In this sense, when thinking about a system of cities, cities are not independent and isolated systems. The correlation between the different processes inside cities needs to be untangled \cite{curiel_heartbeat_2021}, in addition to considering their co-evolution \cite{pumain_evolutionary_2006}.

In a paper in this special issue \cite{altmann_spatial_2020}, the authors considered proximity and interactions between cities, and found that the value of the exponent changes under such considerations. In \cite{ribeiro_association_2021}, the authors looked at 96 countries and found that the effect of urban scaling of GDP is correlated to the population distribution. On the other hand, the authors in \cite{bergs_spatial_2021} showed that the observed Zipf’s law emerges from the autocorrelation of the distribution of cities. These latest works emphasize the increasing importance of integrating interactions between cities and correlations into any theory or modelling of cities. The importance of transport as an enabler of interactions, shaping the distribution of functions and evolution of the form of the city, has been pointed out for over a century ago. Nowadays, many of the interactions are channelled over the internet, making the physical distance obsolete for some of the processes. Understanding the extent of the impact of new technologies, in particular of those as widespread as this one, is essential. The authors in \cite{tranos_ubiquitous_2021} contribute to this discussion, by analysing the impact of information and communication technologies (ICT) on the agglomeration benefits, and by pointing out its effect on the distribution of city size.

\section*{Morphology and inequality}

The previous section mentioned the importance of transport in facilitating and enabling interactions and different types of flows between the system’s components. In the next section we will look in more detail at some of the models of flows. What is important to note at this stage, is that the spatial patterns observed in cities are the outcome of years of restructuring, where positive feedbacks might have reinforced certain paths. Such a reinforcement is the outcome of bottom-up self-organising processes and top-down interventions. These, however, do not necessarily lead to more resilient, nor optimal cities. 
The \emph{form} of a city can hence be seen as the outcome of all the above-mentioned processes, where no unique solution can be defined. The quest of identifying the interplay between \emph{form} and \emph{function} has been ongoing for many decades now: does form follow function or does function adapt to form? Their co-evolution is a complex mechanism involving slow and fast dynamics.  In order to couple these systems, \emph{form} needs to be clearly characterised. This has led to a wide variety of frameworks being introduced to properly define the morphology of cities. 

On the one hand, this can be thought in terms of the shape of its built components, such as the plots and buildings constituting cities \cite{steadman_classification_2000,marshall_streets_2004}. This cannot be disentangled from the \emph{age} of the buildings, since the introduction of new buildings into the city is very much dependent on the probability of buildings being demolished and plots being repurposed \cite{noauthor_colouring_nodate,uhl_century_2021}. 
As cities evolve, some of the reinforced patterns correspond many times to negative characteristics, such as segregation and poverty \cite{vaughan_space_2005}. Understanding the effect of the physical form on segregation is an urgent and important problem to address.
On the other hand, other reinforcement processes have left a physical imprint corresponding to the street network.
And although these are the outcome of different intervening constraints around the world: socio-political, historical, and geographical; the emergent pattern can universally be recognised as a fractal \cite{batty_fractal_1994,frankhauser_fractal_1998}. Furthermore, through the advancement of network science \cite{newman_networks_2018,barabasi_scale-free_2009}, street networks have been analysed through the application and development of different centrality measures \cite{rosvall_networks_2005,porta_network_2006,porta_network_2006P2,barthelemy_self-organization_2013,strano_elementary_2012,jiang_topological_2004,boeing_osmnx_2017,molinero_angular_2017}. In parallel, a whole discipline emerged from the effort of connecting the form of cities through network-like approaches to its function: \emph{space syntax} \cite{hillier_social_1984}. 

With the advent of sophisticated computational methods allowing for the collection, manipulation, and classification of big datasets, a characterisation of cities at a large scale has been possible \cite{louf_typology_2014}, in particular making use of methods from machine learning \cite{kempinska_modelling_2019,milojevic-dupont_learning_2020,simons_untangling_2021}. These methods have been refined through the combination of different datasets, including remote sensing data, such as LiDAR \cite{bonczak_large-scale_2019,huo_supervised_2018,noauthor_mit_2021}.

As paths got reinforced and cities evolved, the process took place in a non-uniform way, leading to different growth rates in the system. This produced a multifractal organisation of the street network \cite{murcio_multifractal_2015, long_multifractal_2021}, and of urban systems in general \cite{thomas_clustering_2010}. The self-organised processes, together with top-down interventions are responsible of the observed morphologies of cities. London for example, evolved from a multifractal to a mono-fractal, following the introduction of a greenbelt around the city constraining its growth \cite{murcio_multifractal_2015}. In more extreme cases, interventions such as the one undertaken by Hausmann in Paris in the late 19th century, have seen large parts of the system destroyed to reconstruct a new order.

The morphology of a city constraints the spatial distribution of functions. Multifractal methods can be used to understand both, the spatial disparities, and the skewness of the distribution \cite{salat_multifractal_2017,salat_uncovering_2018} leading to inequality. 
On the other hand, there are proposals on reducing inequality through the redistribution of flows \cite{louail_crowdsourcing_2017}. The topics of inequality and segregation could have their own special issue in the journal, and it is not our intention to propose a review of the topics here. It is important nevertheless, to mention their embeddedness within many different layers of the city, including mobility patterns \cite{pepe_covid-19_2020,moro_mobility_2021}. 
This allows us to further motivate the role of flows within urban systems as an essential component. For example, these play a central role in the characterisation of spaces \cite{botta_modelling_2021,alhazzani_urban_2021}, and in defining their importance within the system.   

\section*{Hierarchical organisation}

The previous section revealed that the morphology observed in cities is a consequence of the reinforcement of the connectivity between places occurring in a non-homogeneous way. 
The heterogeneous interactions take place at different scales. For example, at a very granular level, denser parts of the city have a higher probability of encounters, and these could be identified as neighbourhoods. At the next level, cities can be represented in terms of their neighbourhoods as nodes, which are many times defined in terms of administrative boundaries, such as boroughs or census tracts. Neighbourhoods also present different degrees of interaction between them, which can be characterised in terms of commuting flows \cite{batty_visualizing_2018}, or any other type of interaction proxy \cite{cottineau_nested_2020}. This generates a nested structure which encodes a hierarchical structure where different parts of the system are more connected than others, generating feedback loops across scales. Such a structure is not confined to cities; interactions can be defined between cities, regions, countries, or between the different scales \cite{arcaute_cities_2016}. Furthermore, hierarchical organisation is a commonly observed pattern in complex systems \cite{pumain_hierarchy_2006,self-organization_2003}. 

At this point, it is important to recall that many of the observed processes are the outcome of interactions which have been modulated by the their spatial distributions and the speed of transportation \cite{pumain_evolutionary_2018}. Hence, the co-evolution of different systems within cities is coupled to technological advancements. And although many of the interactions take place in the cyberspace, mobility in cities also shows a strong hierarchical organisation, see for instance \cite{bassolas_2019_hierarchy,alessandretti_scales_2020,arcaute_hierarchies_2020}. Such organisation mimics Christaller's idea in the central place theory for the hierarchy between cities, with some areas being of higher category than others and attracting or emitting most of the trips (hotspots) \cite{louail_2014_mobile,louail_2015_uncovering}. This is a mesoscopic level of description of the city’s organisation, since it requires the analysis of aggregated mobility and of the structure of the hotspots and their levels in space. However, as was shown in \cite{bassolas_2019_hierarchy}, the fact that a city is widespread or more compact in terms of its centres of activity can be connected to indicators of the quality of life such as the transport modes use to travel to work, the levels of pollution and public health records.  

\section*{Urban mobility }

The role of mobility in cities is to interconnect the different zones constituting an urban area. There are plenty of reasons why people residing at a certain neighbourhood need to travel like, for instance, to work, to find goods and services, for leisure purposes, etc. As shown by the last results discussed on the hierarchical organisation of cities, both morphological aspects and mobility are strongly entangled. Nevertheless, these two questions have been addressed in the literature as two separate issues. The next important quest is to bring these two fields together. In the meantime, let us look at the work done on mobility, in particular at the data needed to characterise it and at the models proposed to explain it at different scales.   

Traditionally, census surveys collect the present residence location of citizens and, in many countries they include a question on the place of residence in the previous census. Since the period between surveys is around 10 years, this information provided a basis for the very first analyses on migration flows \cite{ravenstein_1885_laws}. Much later, already in the 2000s for the US for example, the surveys incorporated the question of county of work. In this way, it was possible to outline the commuting mobility flows at country scale. In terms of surveys, there have been as well a tradition on performing local travel surveys in several cities. The final goal is to improve the management of the public transport system, but since the questions asked are more specific these are very valuable information sources. Unfortunately, in most of the cases transport surveys are not standardized across cities and they have an eminently local character. 

For digital data sources, one of the first works was related to the "where is George?" experiment \cite{brockmann_2006_scaling} in which individuals introduced in a web the code of the bank notes across the US. Following the notes through their locations, it was possible to gain insights on the potential travels. However, this could only provide a type of proxy of mobility. Mobile phone records introduced a much more direct way to measure it. The first results focused on land use in different areas of the cities \cite{ratti_2006_mobile,calabrese_2006_real,soto_2011_automated,toole_2012_inferring}, while a later work by M. Gonz\'alez et al. analysed individual mobility patterns \cite{gonzalez_2008_understanding}. After that, there has been plenty of work with this type of data, including the analysis of the social network in space, mobility in cities \cite{lenormand_2014_cross}, etc (see, for instance, \cite{Blondel_2015_CDRreview} for a review). Other digital sources of data include online social networks as Twitter or Facebook.

\section*{Models of mobility }

Theoretical models must adjust to the scales that they intend to reproduce. Mobility can be seen as a personal phenomenon in which individuals or agents crawl across the city, or as an aggregated entity connecting city zones with flows of people travelling between them. The approaches used to address these two scales are based on different levels of knowledge on the system and input data details. While individual-based models attempt to reproduce trajectories, something that requires rules on people decision-making and data on their trip demands, aggregated models need only parameters accounting for the city-zones properties as sources or sinks for travels.

\section*{Individual-based models}

The original individual models were based on concepts developed for random walks \cite{einstein_1905_movement} and L\'evy flights (see for instance \cite{barbosa_2018_human} for a recent review). The main variable is the position $X_i(t)$ of every agent $i$ at time $t$ and the metrics are built out of it as, for instance, the mean square displacement per agent or the mean (median) radius of gyration \cite{gonzalez_2008_understanding}. The basic versions of these models lead to diffusion of different types depending on the particular statistics of the trip-lengths or jumps. Nevertheless, the population spreads over all the space and the final state tends to have uniform density, which is not a very realistic feature in urban systems. 

More elaborated theoretical frameworks add aspects such as continuous time movements to the random walks, so that agents can travel at any time and do not do it in a synchronised way, or fractional random walks in which the next jump is a product of a process with long memory with respect to the previous displacements. As occurs with L\'evy flights, it is possible to consider ambivalent models in which both the space jumps and the waiting times follow power-laws. These ideas are motivated by empirical observations in different datasets ~\cite{brockmann_2006_scaling, gonzalez_2008_understanding, zhao_2008_empirical, song_2010_modelling}.

One issue important to mention is that individuals typically return to one or several locations, e.g. home or work. Taking as basis the previous random-walk-like frameworks, models have been proposed to include return to previous visited locations \cite{gillis_1970_expected,song_2010_modelling}. Usually, the probability of returning is proportional to the number of visits that an individual has paid to a place, which defines the so-called preferential return \cite{song_2010_modelling}. The rest of the time the agents explore new environments. It has been also proposed that individuals' interest in a place may also decline with time in some cases. This mechanism helps to modify the center of mass of each agent's movements after a medium-long period of time and goes under the name of recency \cite{barbosa_2015_effect}.   Other studies have shown that people visit regularly a finite number of places, and if a new place becomes frequent, another one is abandoned \cite{alessandretti_evidence_2018}.

Humans are a social animal, and, therefore, realistic models need to contemplate the effect of social interactions on mobility \cite{axhausen_2005_social,carrasco_2006_exploring,dugunji_2005_discrete}. There are plenty of circumstances in which this becomes an important question: people may travel in family or in other generic groups; additionally, they may share a common destination synchronising their trajectories to meet somewhere. From a data point of view, traces of these behaviours can be observed using online social networks \cite{liben-nowell_2005_geographic,carrasco_2008_collecting,carrasco_2008_agency}, cell phone records \cite{lambiotte_2008_geographical,krings_2009_urban,phithakkitnukoon_2012_socio}  and surveys \cite{carrasco_2008_how,vandenberg_2013_path}. This phenomenon can help to improve individual model predictions on trajectories thanks to the correlations in displacements with the social contacts \cite{de_2013_interdependence}.  The effect of group mobility has been considered in transportation microsimulations since a little more than a decade ago  \cite{axhausen_2005_social,carrasco_2006_exploring,dugunji_2005_discrete,paez_2007_social,arentze_2008_social,carrasco_2009_social,hackney_2011_coupled,ronald_2012_modeling,sharmeen_2014_dynamics}. Interestingly, the relation between social network and mobility is bidirectional. Our friends determine some of our mobility patterns, but we establish as well new social relations with people with meet in the visited locations. Traces of the organisation of mobility can be detected in the social network, which allowed for the definition of models characterising this interplay  \cite{gonzalez_2006_system, grabowicz_2014_entangling,toole_2015_coupling}.

\section*{Modelling aggregated mobility flows}

Passing now to the aggregated models, generally trip flows are characterized as origin-destination (OD) matrices. Every element of the matrix at row $i$ and column $j$ conveys the information on the number of trips between areas $i$ (origin) and $j$ (destination). An OD matrix can also be represented as a directed weighted network, with links pointing from the area of origin to the one of destination and the weight standing for the number of trips. These matrices have been used traditionally to express the trip demand between zones of a city and are, therefore, an essential tool for infrastructure planning. Finding models able to infer the OD matrices from non-mobility variables is thus a question of great relevance. Traditionally, two family of approaches have dominated these endeavours: the intervening opportunities and the gravity models.

Intervening opportunities models were initially introduced in 1940 \cite{stouffer_1940_intervening}. The idea guiding these models is that the population behaves as a source of trips and the destination depends on the number of opportunities an agent sees around her/his residential area. The probability of the agent to select one of these opportunities and, therefore, to set one destination for her trips relies on different functions that try to quantify the quality of the opportunity. These models have been profusely studied for a long time, see for example Refs. \cite{heanus_1966_comparative, ruiter_1967_toward, haynes_1973_intermetropolitan, wilson_1970_urban, fik_1990_spatial, akwawua_2001_development}. Recently, a self-consistent version called radiation model has been introduced \cite{simini_2012_universal}. The radiation model considers the "quality"  of opportunities as a random variable and, as a consequence, the one to be selected should correspond to the largest quality value. The selection of extremes under these circumstances generates a few families of universal distributions, depending on the nature of the original random variable and its moments. This universality allows thus to close the expressions and find an equation for the flows of trips between areas. Later, other versions of the model have been considered to improve the treatment of the spatial scales and the nonlinear relation between opportunities quality and zone attractiveness \cite{simini_2013_human, yang_2014_limits,carra_2016_modeling}.

The gravity model \cite{zipf_1946_p1} takes the population of the origin as the source for the trips, so the trip number is proportional to it. The attractiveness of the destination area is related as well to its population, the relation can be linear in the simplest form of the model or, more generically, nonlinear. But the main question differentiating the gravity model from the intervening opportunities ones is that the flow of trips decays with the distance between origin and destination with a deterrence function. Most commonly, such a function can be an exponential or a power-law and its form may depend on the geographical scale considered, the purpose of the mobility or the transportation mode \cite{barthelemy_2011_spatial}. The simplicity of this law has made it very popular for applications, for instance, in transport infrastructure planning \cite{erlander_1990_gravity,ortuzar_2011_modeling}, geography \cite{wilson_1970_urban} and spatial economics \cite{karemera_2000_gravity,patuelli_2007_network}. The gravity model can be deduced from a maximum entropy principle \cite{wilson_1970_entropy}. It is also important to mention that the basic equation for the gravity model is unconstrained: given the population and distance between areas, one obtains directly an estimation of the flows. This is not the case for the radiation model, which is origin contained with the number of outgoing trips per area given as an input. 
Constrained can also be considered in gravity models, these can be at the origin, at the destination or at both \cite{lenormand_2015_systematic,wilson_family_1971}.
Assessing the effectiveness of these models is ongoing research \cite{hilton_predictive_2020}, in some cases the scale of the system plays an important role \cite{masucci_gravity_2013}, while in some others, the results from the radiation and gravity models are in agreement when assessing interventions \cite{piovani_measuring_2018}.

\section*{Comparison between models and future trends}

In the case of commuting, there have been several works to compare the performance of both families of models.  For example, the flows predicted by models with different levels of constraints  are directly compared with the empirical values in Ref.  \cite{lenormand_2015_systematic}. The results seem to favour the exponential gravity model, even though by a narrow margin. More recently, a field theoretical framework for mobility has been proposed \cite{mazzoli_2019_field}. In this case, the average mobility of the flows out of each area are vectorially averaged and the different models have been used to explain the empirical patterns. Again, for the case of commuting the winner was a gravity model with exponential deterrence function, and in this case the margin with the other models was much wider. 

 Hierarchy  in space also emerges naturally if one thinks about how the areas are embedded into one another: neighbourhoods form part of a city, cities lie within regions, regions constitute countries, etc. 
 Recently, a model to exploit this hierarchical organisation has been proposed, where mobility at different scales gives rise to a nested structure of containers \cite{alessandretti_scales_2020}. 
 Finally, and recalling the Song's model of preferential return, another model has been introduced exploring the role of the frequency of visits of individuals and mobility patterns, giving rise to a scaling law between the number of visitors and the product of the visiting frequency and the travel distance \cite{schlapfer_2021_trip}.



\section*{Conclusions}

In the quest to model processes within cities, whole disciplines have emerged over the last century, from economics to transport modelling. Although each of these has produced great advancements, the time has come to couple them. 
Firstly, we discussed the need to include interactions and correlations within and between cities when accounting for agglomeration effects. We pointed out at the role of transport as an enabler of the interactions driving many of the observed processes. 
These take place in space, and the morphology of a city has proven to play a substantial role in reinforcing certain patterns determining the location of functions. Furthermore, such patterns present a hierarchical organisation.

After this, we have discussed mobility, a very interrelated phenomenon with morphology, and that encompasses the interactions between places. We introduced the two main modelling approaches to mobility: individual-based models, and models of aggregated mobility flows. 
The past decade has seen a surge in important works on mobility, driven by an increased availability of mobility data through mobile phones and other social media data, which has helped fine tune the models.

Agglomeration effects cannot be disconnected from the location of functions, nor from the differentiated opportunities given by transport, which are manifested through the mobility patterns left by people in a city. In this sense, the time is now ripe to try to integrate all these different components towards a better understanding of cities.

\section*{Acknowledgments}

E.A. acknowledges funding from the UK Engineering and Physical Sciences Research Council (EPSRC) (Grant No. EP/M023583/1).
J.J.R. acknowledges funding from the project PACSS (RTI2018-093732-B-C22) of the MCIN/AEI/10.13039/501100011033/ and by EU through FEDER funds (A way to make Europe), and also from the Maria de Maeztu program MDM-2017-0711 of the MCIN/AEI/10.13039/501100011033/. 

\section*{Author contributions}

E.A. and J.J.R. contributed equally in conceiving and writing the paper.

\section*{Competing Interests statement}
The authors declare no conflict of interest.


\begin{thebibliography}{100}

\bibitem{christaller_central_1933}
Christaller W.
\newblock Central Places in Southern Germany.
\newblock Englewood Cliffs, {NJ}: Prentice-Hall, 1966. Original work published
  in 1933 as "Die Zentrale Orte in Suddeutschalnd", Jena, Germany: Gustav
  Fisher; 1933.

\bibitem{losch_economics_1954}
L{\"o}sch A.
\newblock The economics of location.
\newblock New Haven; 1954.

\bibitem{Haggett_Chorley1969}
Haggett P, Chorley R.
\newblock Network analysis in geography.
\newblock Edward Arnold; 1969.

\bibitem{berry_cities_1964}
Berry BJL.
\newblock Cities as {S}ystems Within {S}ystems of {C}ities.
\newblock Papers in Regional Science. 1964;13(1):147--163.
\newblock Available from:
  \url{https://rsaiconnect.onlinelibrary.wiley.com/doi/abs/10.1111/j.1435-5597.1964.tb01283.x}.

\bibitem{pumain_pour_1997}
Pumain D.
\newblock Pour une th{\'e}orie {\'e}volutive des villes.
\newblock L'Espace g{\'e}ographique. 1997;26(2):119--134.
\newblock Available from: \url{https://www.jstor.org/stable/44381391}.

\bibitem{pumain_les_1992}
Pumain D.
\newblock Les syst{\`e}mes de villes.
\newblock Encyclop{\'e}die de G{\'e}ographie. 1992;p.~20.

\bibitem{batty_new_2013}
Batty M.
\newblock The New Science of Cities.
\newblock {MIT} Press; 2013.

\bibitem{batty_inventing_2018}
Batty M.
\newblock Inventing {Future} {Cities}.
\newblock Cambridge, MA, USA: MIT Press; 2018.

\bibitem{marshall_a._principles_1890}
{Marshall, A }.
\newblock Principles of Economics.
\newblock Macmillan and Co.; 1890.
\newblock Available from:
  \url{http://archive.org/details/principlesecono00marsgoog}.

\bibitem{duranton_2004_micro}
Duranton G, Puga D.
\newblock Micro-foundations of urban agglomeration economies.
\newblock In: Handbook of regional and urban economics. vol.~4. Elsevier; 2004.
  p. 2063--2117.

\bibitem{glaeser_introduction_2010}
Glaeser EL.
\newblock {Agglomeration Economics}.
\newblock University of Chicago Press; 2010.

\bibitem{bettencourt_growth_2007}
Bettencourt LMA, Lobo J, Helbing D, K{\"u}hnert C, West GB.
\newblock Growth, innovation, scaling, and the pace of life in cities.
\newblock Proc Natl Acad Sci {USA}. 2007;104(17):7301--7306.
\newblock Available from:
  \url{http://www.pnas.org/content/104/17/7301.abstract}.

\bibitem{pumain_scaling_2004}
Pumain D.
\newblock Scaling Laws and Urban Systems.
\newblock {SFI} Working Paper 2004-02-002. 2004;Available from:
  \url{https://www.santafe.edu/research/results/working-papers/scaling-laws-and-urban-systems}.

\bibitem{pumain_evolutionary_2006}
Pumain D, Paulus F, Vacchiani-Marcuzzo C, Lobo J.
\newblock An evolutionary theory for interpreting urban scaling laws.
\newblock Cybergeo. 2006;343.
\newblock Available from: \url{DOI:10.1068/b32045}.

\bibitem{strano_rich_2016}
Strano E, Sood V.
\newblock Rich and Poor Cities in Europe. An Urban Scaling Approach to Mapping
  the European Economic Transition.
\newblock {PLOS} {ONE}. 2016;11(8):e0159465.
\newblock Available from:
  \url{https://journals.plos.org/plosone/article?id=10.1371/journal.pone.0159465}.

\bibitem{samaniego_cities_2008}
Samaniego H, Moses ME.
\newblock Cities as {Organisms}: {Allometric} {Scaling} of {Urban} {Road}
  {Networks}.
\newblock Journal of Transport and Land Use. 2008;1(1).
\newblock Available from:
  \url{https://www.jtlu.org/index.php/jtlu/article/view/29}.

\bibitem{gomez-lievano_statistics_2012}
Gomez-Lievano A, Youn H, Bettencourt LMA.
\newblock The Statistics of Urban Scaling and Their Connection to Zipf's Law.
\newblock {PLoS} {ONE}. 2012;7(7):e40393.

\bibitem{ribeiro_model_2017}
Ribeiro FL, Meirelles J, Ferreira FF, Neto CR.
\newblock A model of urban scaling laws based on distance dependent
  interactions.
\newblock Royal Society Open Science. 2017;4(3):160926.
\newblock Available from:
  \url{https://royalsocietypublishing.org/doi/full/10.1098/rsos.160926}.

\bibitem{meirelles_evolution_2018}
Meirelles J, Neto CR, Ferreira FF, Ribeiro FL, Binder CR.
\newblock Evolution of urban scaling: Evidence from Brazil.
\newblock {PLOS} {ONE}. 2018;13(10):e0204574.
\newblock Available from:
  \url{https://journals.plos.org/plosone/article?id=10.1371/journal.pone.0204574}.

\bibitem{cabrera-arnau_effect_2020}
Cabrera-Arnau C, Bishop SR.
\newblock The effect of dragon-kings on the estimation of scaling law
  parameters.
\newblock Sci Rep. 2020;10(1):20226.
\newblock Available from:
  \url{https://www.nature.com/articles/s41598-020-77232-6}.

\bibitem{oliveira_large_2014}
Oliveira EA, Andrade JS, Makse HA.
\newblock Large cities are less green.
\newblock Scientific Reports. 2014;4(4235).

\bibitem{fragkias_does_2013}
Fragkias M, Lobo J, Strumsky D, Seto KC.
\newblock Does Size Matter? Scaling of {CO}2 Emissions and U.S. Urban Areas.
\newblock {PLoS} {ONE}. 2013;8(6):e64727.

\bibitem{fuller_scaling_2009}
Fuller RA, Gaston KJ.
\newblock The scaling of green space coverage in European cities.
\newblock Biology Letters. 2009;5(3):352--355.
\newblock Available from:
  \url{http://rsbl.royalsocietypublishing.org/content/5/3/352.abstract}.

\bibitem{louf_how_2014}
Louf R, Barthelemy M.
\newblock How congestion shapes cities: from mobility patterns to scaling.
\newblock Scientific Reports. 2014;4:5561.
\newblock Available from: \url{https://www.nature.com/articles/srep05561}.

\bibitem{depersin_global_2018}
Depersin J, Barthelemy M.
\newblock From global scaling to the dynamics of individual cities.
\newblock {PNAS}. 2018;115(10):2317--2322.
\newblock Available from: \url{https://www.pnas.org/content/115/10/2317}.

\bibitem{sarkar_urban_2018}
Sarkar S.
\newblock Urban scaling and the geographic concentration of inequalities by
  city size.
\newblock Environment and Planning B: Urban Analytics and City Science. 2018;p.
  2399808318766070.
\newblock Available from: \url{https://doi.org/10.1177/2399808318766070}.

\bibitem{arcaute_constructing_2015}
Arcaute E, Hatna E, Ferguson P, Youn H, Johansson A, Batty M.
\newblock Constructing cities, deconstructing scaling laws.
\newblock Journal of The Royal Society Interface. 2015;12(102):20140745.
\newblock Available from:
  \url{https://royalsocietypublishing.org/doi/full/10.1098/rsif.2014.0745}.

\bibitem{leitao_is_2016}
Leit{\~a}o JC, Miotto JM, Gerlach M, Altmann EG.
\newblock Is this scaling nonlinear?
\newblock Royal Society Open Science. 2016;3(7):150649.
\newblock Available from:
  \url{https://royalsocietypublishing.org/doi/full/10.1098/rsos.150649}.

\bibitem{shalizi_scaling_2011}
Shalizi CR.
\newblock Scaling and {Hierarchy} in {Urban} {Economies}.
\newblock arXiv:11024101 [physics, stat]. 2011;Available from:
  \url{http://arxiv.org/abs/1102.4101}.

\bibitem{gerlach_testing_2019}
Gerlach M, Altmann EG.
\newblock Testing {Statistical} {Laws} in {Complex} {Systems}.
\newblock Phys Rev Lett. 2019;122(16):168301.
\newblock Available from:
  \url{https://link.aps.org/doi/10.1103/PhysRevLett.122.168301}.

\bibitem{corral_truncated_2020}
Corral {\'A}, Udina F, Arcaute E.
\newblock Truncated lognormal distributions and scaling in the size of
  naturally defined population clusters.
\newblock Phys Rev E. 2020;101(4):042312.
\newblock Available from:
  \url{https://link.aps.org/doi/10.1103/PhysRevE.101.042312}.

\bibitem{hong_common_2018P}
Hong I, Frank MR, Rahwan I, Jung WS, Youn H.
\newblock A common trajectory recapitulated by urban economies.
\newblock Science Advances. 2020;6(eaba4934).

\bibitem{ribeiro_relation_2020}
Ribeiro FL, Meirelles J, Netto VM, Neto CR, Baronchelli A.
\newblock On the relation between transversal and longitudinal scaling in
  cities.
\newblock PLOS ONE. 2020;15(5):e0233003.
\newblock Available from:
  \url{https://journals.plos.org/plosone/article?id=10.1371/journal.pone.0233003}.

\bibitem{ortman_cities_2020}
Ortman SG, Lobo J, Smith ME.
\newblock Cities: {Complexity}, theory and history.
\newblock PLOS ONE. 2020;15(12):e0243621.
\newblock Available from:
  \url{https://journals.plos.org/plosone/article?id=10.1371/journal.pone.0243621}.

\bibitem{altmann_spatial_2020}
Altmann EG.
\newblock Spatial interactions in urban scaling laws.
\newblock PLOS ONE. 2020;15(12):e0243390.
\newblock Available from:
  \url{https://journals.plos.org/plosone/article?id=10.1371/journal.pone.0243390}.

\bibitem{ribeiro_association_2021}
Ribeiro HV, Oehlers M, Moreno-Monroy AI, Kropp JP, Rybski D.
\newblock Association between population distribution and urban {GDP} scaling.
\newblock PLOS ONE. 2021;16(1):e0245771.
\newblock Available from:
  \url{https://journals.plos.org/plosone/article?id=10.1371/journal.pone.0245771}.

\bibitem{meakin_fractals_1998}
Meakin P.
\newblock Fractals, Scaling and Growth Far from Equilibrium.
\newblock Cambridge University Press; 1998.

\bibitem{curiel_heartbeat_2021}
Curiel RP, Patino JE, Duque JC, O'Clery N.
\newblock The heartbeat of the city.
\newblock PLOS ONE. 2021;16:e0246714.

\bibitem{bergs_spatial_2021}
Bergs R.
\newblock Spatial dependence in the rank-size distribution of cities: weak but
  not negligible.
\newblock PLOS ONE. 2021;16(2):e0246796.
\newblock Available from:
  \url{https://journals.plos.org/plosone/article?id=10.1371/journal.pone.0246796}.

\bibitem{tranos_ubiquitous_2021}
Tranos E, Ioannides YM.
\newblock Ubiquitous digital technologies and spatial structure; an update.
\newblock PLOS ONE. 2021;16(4):e0248982.
\newblock Available from:
  \url{https://journals.plos.org/plosone/article?id=10.1371/journal.pone.0248982}.

\bibitem{steadman_classification_2000}
Steadman P, Bruhns HR, Holtier S, Gakovic B, Rickaby PA, Brown FE.
\newblock A {Classification} of {Built} {Forms}.
\newblock Environ Plann B Plann Des. 2000;27(1):73--91.
\newblock Available from: \url{https://doi.org/10.1068/bst7}.

\bibitem{marshall_streets_2004}
Marshall S.
\newblock Streets and {Patterns}.
\newblock London: Routledge; 2004.

\bibitem{noauthor_colouring_nodate}
Colouring {London};.
\newblock Available from: \url{https://colouring.london}.

\bibitem{uhl_century_2021}
Uhl JH, Connor DS, Leyk S, Braswell AE.
\newblock A century of decoupling size and structure of urban spaces in the
  {United} {States}.
\newblock Commun Earth Environ. 2021;2(1):1--14.
\newblock Available from:
  \url{https://www.nature.com/articles/s43247-020-00082-7}.

\bibitem{vaughan_space_2005}
Vaughan L, Clark DLC, Sahbaz O, Haklay MM.
\newblock Space and exclusion: does urban morphology play a part in social
  deprivation?
\newblock Area. 2005;37(4):402--412.
\newblock Available from:
  \url{https://onlinelibrary.wiley.com/doi/abs/10.1111/j.1475-4762.2005.00651.x}.

\bibitem{batty_fractal_1994}
Batty M, Longley P.
\newblock Fractal Cities: A Geometry of Form and Function.
\newblock Academic Press, San Diego, {CA} and London; 1994.

\bibitem{frankhauser_fractal_1998}
Frankhauser P.
\newblock The fractal approach. {A} new tool for the spatial analysis of urban
  agglomerations.
\newblock Population: An English Selection. 1998;p. 205--240.

\bibitem{newman_networks_2018}
Newman M.
\newblock Networks.
\newblock Oxford University Press; 2018.

\bibitem{barabasi_scale-free_2009}
Barab{\'a}si AL.
\newblock Scale-{Free} {Networks}: {A} {Decade} and {Beyond}.
\newblock Science. 2009;325(5939):412--413.
\newblock Available from:
  \url{https://www.science.org/doi/full/10.1126/science.1173299}.

\bibitem{rosvall_networks_2005}
Rosvall M, Trusina A, Minnhagen P, Sneppen K.
\newblock Networks and {Cities}: {An} {Information} {Perspective}.
\newblock Phys Rev Lett. 2005;94(2):028701.
\newblock Available from:
  \url{https://link.aps.org/doi/10.1103/PhysRevLett.94.028701}.

\bibitem{porta_network_2006}
Porta S, Crucitti P, Latora V.
\newblock The {Network} {Analysis} of {Urban} {Streets}: {A} {Primal}
  {Approach}.
\newblock Environ Plann B Plann Des. 2006;33(5):705--725.
\newblock Available from: \url{https://doi.org/10.1068/b32045}.

\bibitem{porta_network_2006P2}
Porta S, Crucitti P, Latora V.
\newblock The network analysis of urban streets: {A} dual approach.
\newblock Physica A: Statistical Mechanics and its Applications.
  2006;369(2):853--866.
\newblock Available from:
  \url{https://www.sciencedirect.com/science/article/pii/S0378437106001282}.

\bibitem{barthelemy_self-organization_2013}
Barthelemy M, Bordin P, Berestycki H, Gribaudi M.
\newblock Self-organization versus top-down planning in the evolution of a
  city.
\newblock Sci Rep. 2013;3(1):2153.
\newblock Available from: \url{https://www.nature.com/articles/srep02153}.

\bibitem{strano_elementary_2012}
Strano E, Nicosia V, Latora V, Porta S, Barth{\'e}lemy M.
\newblock Elementary processes governing the evolution of road networks.
\newblock Sci Rep. 2012;2(1):296.
\newblock Available from: \url{https://www.nature.com/articles/srep00296}.

\bibitem{jiang_topological_2004}
Jiang B, Claramunt C.
\newblock Topological Analysis of Urban Street Networks.
\newblock Environ Plann B Plann Des. 2004;31(1):151--162.
\newblock Available from: \url{https://doi.org/10.1068/b306}.

\bibitem{boeing_osmnx_2017}
Boeing G.
\newblock {OSMnx}: {New} methods for acquiring, constructing, analyzing, and
  visualizing complex street networks.
\newblock Computers, Environment and Urban Systems. 2017;65:126--139.
\newblock Available from:
  \url{https://www.sciencedirect.com/science/article/pii/S0198971516303970}.

\bibitem{molinero_angular_2017}
Molinero C, Murcio R, Arcaute E.
\newblock The angular nature of road networks.
\newblock Scientific Reports. 2017;7(1):4312.
\newblock Available from:
  \url{https://www.nature.com/articles/s41598-017-04477-z}.

\bibitem{hillier_social_1984}
Hillier B, Hanson J.
\newblock The {Social} {Logic} of {Space}.
\newblock Cambridge: Cambridge University Press; 1984.
\newblock Available from:
  \url{https://www.cambridge.org/core/books/social-logic-of-space/6B0A078C79A74F0CC615ACD8B250A985}.

\bibitem{louf_typology_2014}
Louf R, Barthelemy M.
\newblock A typology of street patterns.
\newblock Journal of The Royal Society Interface. 2014;11(101):20140924.
\newblock Available from:
  \url{https://royalsocietypublishing.org/doi/10.1098/rsif.2014.0924}.

\bibitem{kempinska_modelling_2019}
Kempinska K, Murcio R.
\newblock Modelling urban networks using {Variational} {Autoencoders}.
\newblock Appl Netw Sci. 2019;4(1):1--11.
\newblock Available from:
  \url{https://appliednetsci.springeropen.com/articles/10.1007/s41109-019-0234-0}.

\bibitem{milojevic-dupont_learning_2020}
Milojevic-Dupont N, Hans N, Kaack LH, Zumwald M, Andrieux F, Soares DdB, et~al.
\newblock Learning from urban form to predict building heights.
\newblock PLOS ONE. 2020;15(12):e0242010.
\newblock Available from:
  \url{https://journals.plos.org/plosone/article?id=10.1371/journal.pone.0242010}.

\bibitem{simons_untangling_2021}
Simons GD.
\newblock Untangling urban data signatures: unsupervised machine learning
  methods for the detection of urban archetypes at the pedestrian scale.
\newblock arXiv:210615363 [physics]. 2021;Available from:
  \url{http://arxiv.org/abs/2106.15363}.

\bibitem{bonczak_large-scale_2019}
Bonczak B, Kontokosta CE.
\newblock Large-scale parameterization of {3D} building morphology in complex
  urban landscapes using aerial {LiDAR} and city administrative data.
\newblock Computers, Environment and Urban Systems. 2019;73:126--142.
\newblock Available from:
  \url{https://www.sciencedirect.com/science/article/pii/S0198971518300176}.

\bibitem{huo_supervised_2018}
Huo LZ, Silva CA, Klauberg C, Mohan M, Zhao LJ, Tang P, et~al.
\newblock Supervised spatial classification of multispectral {LiDAR} data in
  urban areas.
\newblock PLOS ONE. 2018;13(10):e0206185.
\newblock Available from:
  \url{https://journals.plos.org/plosone/article?id=10.1371/journal.pone.0206185}.

\bibitem{noauthor_mit_2021}
{MIT} senseable city lab analyzes brazilian favela's {3D} morphology utilizing
  {LiDAR}; 2021.
\newblock Available from:
  \url{https://www.designboom.com/technology/mit-senseable-city-lab-brazilian-favelas-3d-morphology-lidar-04-26-2021/}.

\bibitem{murcio_multifractal_2015}
Murcio R, Masucci AP, Arcaute E, Batty M.
\newblock Multifractal to monofractal evolution of the {London} street network.
\newblock Phys Rev E Stat Nonlin Soft Matter Phys. 2015;92(6):062130.

\bibitem{long_multifractal_2021}
Long Y, Chen Y.
\newblock Multifractal scaling analyses of urban street network structure:
  {The} cases of twelve megacities in {China}.
\newblock PLOS ONE. 2021;16(2):e0246925.
\newblock Available from:
  \url{https://journals.plos.org/plosone/article?id=10.1371/journal.pone.0246925}.

\bibitem{thomas_clustering_2010}
Thomas I, Frankhauser P, Frenay B, Verleysen M.
\newblock Clustering {Patterns} of {Urban} {Built}-up {Areas} with {Curves} of
  {Fractal} {Scaling} {Behaviour}.
\newblock Environ Plann B Plann Des. 2010;37(5):942--954.
\newblock Available from:
  \url{https://journals.sagepub.com/doi/abs/10.1068/b36039}.

\bibitem{salat_multifractal_2017}
Salat H, Murcio R, Arcaute E.
\newblock Multifractal methodology.
\newblock Physica A: Statistical Mechanics and its Applications.
  2017;473:467--487.
\newblock Available from:
  \url{//www.sciencedirect.com/science/article/pii/S0378437117300341}.

\bibitem{salat_uncovering_2018}
Salat H, Murcio R, Yano K, Arcaute E.
\newblock Uncovering inequality through multifractality of land prices: 1912
  and contemporary {Kyoto}.
\newblock PLOS ONE. 2018;13(4):e0196737.
\newblock Available from:
  \url{https://journals.plos.org/plosone/article?id=10.1371/journal.pone.0196737}.

\bibitem{louail_crowdsourcing_2017}
Louail T, Lenormand M, Murillo~Arias J, Ramasco JJ.
\newblock Crowdsourcing the {Robin} {Hood} effect in cities.
\newblock Appl Netw Sci. 2017;2(1):1--13.
\newblock Available from:
  \url{https://appliednetsci.springeropen.com/articles/10.1007/s41109-017-0026-3}.

\bibitem{pepe_covid-19_2020}
Pepe E, Bajardi P, Gauvin L, Privitera F, Lake B, Cattuto C, et~al.
\newblock {COVID}-19 outbreak response, a dataset to assess mobility changes in
  {Italy} following national lockdown.
\newblock Sci Data. 2020 Jul;7(1):230.
\newblock Available from:
  \url{https://www.nature.com/articles/s41597-020-00575-2}.

\bibitem{moro_mobility_2021}
Moro E, Calacci D, Dong X, Pentland A.
\newblock Mobility patterns are associated with experienced income segregation
  in large {US} cities.
\newblock Nat Commun. 2021 Jul;12(1):4633.
\newblock Available from:
  \url{https://www.nature.com/articles/s41467-021-24899-8}.

\bibitem{botta_modelling_2021}
Botta F, Guti{\'e}rrez-Roig M.
\newblock Modelling urban vibrancy with mobile phone and {OpenStreetMap} data.
\newblock PLOS ONE. 2021;16(6):e0252015.
\newblock Available from:
  \url{https://journals.plos.org/plosone/article?id=10.1371/journal.pone.0252015}.

\bibitem{alhazzani_urban_2021}
Alhazzani M, Alhasoun F, Alawwad Z, Gonz{\'a}lez MC.
\newblock Urban attractors: {Discovering} patterns in regions of attraction in
  cities.
\newblock PLOS ONE. 2021;16(4):e0250204.
\newblock Available from:
  \url{https://journals.plos.org/plosone/article?id=10.1371/journal.pone.0250204}.

\bibitem{batty_visualizing_2018}
Batty M.
\newblock Visualizing aggregate movement in cities.
\newblock Philosophical Transactions of the Royal Society B: Biological
  Sciences. 2018;373(1753):20170236.
\newblock Available from:
  \url{https://royalsocietypublishing.org/doi/10.1098/rstb.2017.0236}.

\bibitem{cottineau_nested_2020}
Cottineau C, Arcaute E.
\newblock The nested structure of urban business clusters.
\newblock Appl Netw Sci. 2020;5(1):2.
\newblock Available from: \url{https://doi.org/10.1007/s41109-019-0246-9}.

\bibitem{arcaute_cities_2016}
Arcaute E, Molinero C, Hatna E, Murcio R, Vargas-Ruiz C, Masucci AP, et~al.
\newblock Cities and {Regions} in {Britain} through hierarchical percolation.
\newblock J R Soc Open Science. 2016;3(4).

\bibitem{pumain_hierarchy_2006}
Pumain D, editor. Hierarchy in natural and social sciences.
\newblock Springer; 2006.

\bibitem{self-organization_2003}
Camazine S, Deneubourg JL, Franks NR, Sneyd J, Theraula G, Bonabeau E.
\newblock Self-{Organization} in {Biological} {Systems}; 2003.
\newblock Available from:
  \url{https://press.princeton.edu/books/paperback/9780691116242/self-organization-in-biological-systems}.

\bibitem{pumain_evolutionary_2018}
Pumain D.
\newblock An {Evolutionary} {Theory} of {Urban} {Systems}.
\newblock In: Rozenblat C, Pumain D, Velasquez E, editors. International and
  {Transnational} {Perspectives} on {Urban} {Systems}. Advances in
  {Geographical} and {Environmental} {Sciences}. Singapore: Springer; 2018. p.
  3--18.
\newblock Available from: \url{https://doi.org/10.1007/978-981-10-7799-9_1}.

\bibitem{bassolas_2019_hierarchy}
Bassolas A, Barbosa-Filho H, Dickinson B, Dotiwalla X, Eastham P, Gallotti R,
  et~al.
\newblock Hierarchical organization of urban mobility and its connection with
  city livability.
\newblock Nature communications. 2019;10(1):4817.

\bibitem{alessandretti_scales_2020}
Alessandretti L, Aslak U, Lehmann S.
\newblock The scales of human mobility.
\newblock Nature. 2020;587(7834):402--407.
\newblock Available from:
  \url{https://www.nature.com/articles/s41586-020-2909-1}.

\bibitem{arcaute_hierarchies_2020}
Arcaute E.
\newblock Hierarchies defined through human mobility.
\newblock Nature. 2020;587(7834):372--373.
\newblock Available from:
  \url{https://www.nature.com/articles/d41586-020-03197-1}.

\bibitem{louail_2014_mobile}
Louail T, Lenormand M, {Cantu Ros} OG, Picornell M, Herranz R, Frias-Martinez
  E, et~al.
\newblock From mobile phone data to the spatial structure of cities.
\newblock Scientific Reports. 2014;4.

\bibitem{louail_2015_uncovering}
Louail T, Lenormand M, Picornell M, {Garc{\'\i}a Cant{\'u}} O, Herranz R,
  Frias-Martinez E, et~al.
\newblock Uncovering the spatial structure of mobility networks.
\newblock Nature Communications. 2015;6:6007.

\bibitem{ravenstein_1885_laws}
Ravenstein EG.
\newblock The laws of migration.
\newblock Journal of the Statistical Society of London. 1885;48:167--235.

\bibitem{brockmann_2006_scaling}
Brockmann D, Hufnagel L, Geisel T.
\newblock The scaling laws of human travel.
\newblock Nature. 2006;439(7075):462--465.

\bibitem{ratti_2006_mobile}
Ratti C, Frenchman D, Pulselli RM, Williams S.
\newblock Mobile {{Landscapes}}: {{Using Location Data}} from {{Cell Phones}}
  for {{Urban Analysis}}.
\newblock Environment and Planning B: Planning and Design. 2006;33(5):727--748.

\bibitem{calabrese_2006_real}
Calabrese F, Ratti C.
\newblock Real time rome.
\newblock Networks and Communication studies. 2006;20(3-4):247--258.

\bibitem{soto_2011_automated}
Soto V, Fr{\'\i}as-Mart{\'\i}nez E.
\newblock Automated land use identification using cell-phone records.
\newblock In: Proceedings of the 3rd ACM international workshop on MobiArch;
  2011. p. 17--22.

\bibitem{toole_2012_inferring}
Toole JL, Ulm M, Gonz{\'a}lez MC, Bauer D.
\newblock Inferring land use from mobile phone activity.
\newblock In: Proceedings of the ACM SIGKDD international workshop on urban
  computing; 2012. p. 1--8.

\bibitem{gonzalez_2008_understanding}
Gonzalez MC, Hidalgo CA, Barabasi AL.
\newblock Understanding individual human mobility patterns.
\newblock Nature. 2008;453(7196):779--782.

\bibitem{lenormand_2014_cross}
{Lenormand} M, {Picornell} M, {Cant{\'u}-Ros} OG, {Tugores} A, {Louail} T,
  {Herranz} R, et~al.
\newblock Cross-{{Checking Different Sources}} of {{Mobility Information}}.
\newblock PLoS ONE. 2014;9(8):e105184.

\bibitem{Blondel_2015_CDRreview}
Blondel VD, Decuyper A, Krings G.
\newblock A survey of results on mobile phone datasets analysis.
\newblock EPJ Data Science. 2015;4:10.

\bibitem{einstein_1905_movement}
Einstein A.
\newblock \"Uber die von der molekularkinetischen Theorie der W{\"a}rme
  geforderte Bewegung von in ruhenden Fl{\"u}ssigkeiten suspendierten Teilchen
  (On the movement of small particles suspended in a stationary liquid demanded
  by the molecular-kinetic theory of heat).
\newblock Annalen der physik. 1905;17:549--560.

\bibitem{barbosa_2018_human}
Barbosa H, Barthelemy M, Ghoshal G, James CR, Lenormand M, Louail T, et~al.
\newblock Human mobility: Models and applications.
\newblock Physics Reports. 2018;734:1--74.

\bibitem{zhao_2008_empirical}
Zhao M, Mason L, Wang W.
\newblock Empirical study on human mobility for mobile wireless networks.
\newblock In: Military Communications Conference, 2008. MILCOM 2008. IEEE.
  IEEE; 2008. p. 1--7.

\bibitem{song_2010_modelling}
Song C, Koren T, Wang P, Barab{\'a}si AL.
\newblock Modelling the scaling properties of human mobility.
\newblock Nature Physics. 2010;6(10):818--823.

\bibitem{gillis_1970_expected}
Gillis JE, Weiss GH.
\newblock Expected number of distinct sites visited by a random walk with an
  infinite variance.
\newblock Journal of Mathematical Physics. 1970;11(4):1307--1312.

\bibitem{barbosa_2015_effect}
Barbosa H, de~Lima-Neto FB, Evsukoff A, Menezes R.
\newblock The effect of recency to human mobility.
\newblock EPJ Data Science. 2015;4(1):1--14.

\bibitem{alessandretti_evidence_2018}
Alessandretti L, Sapiezynski P, Sekara V, Lehmann S, Baronchelli A.
\newblock Evidence for a conserved quantity in human mobility.
\newblock Nat Hum Behav. 2018;2(7):485--491.
\newblock Available from:
  \url{https://www.nature.com/articles/s41562-018-0364-x}.

\bibitem{axhausen_2005_social}
Axhausen KW.
\newblock Social networks and travel: some hypotheses.
\newblock In: Donaghy K, Poppelreuter S, Rudinger G, editors. Social dimensions
  of sustainable transport: transatlantic perspectives. London, UK: Ashgate
  Aldershot; 2005. p. 90--108.

\bibitem{carrasco_2006_exploring}
Carrasco JA, Miller EJ.
\newblock Exploring the propensity to perform social activities: social
  networks approach.
\newblock Transportation. 2006;33:463--480.

\bibitem{dugunji_2005_discrete}
Dugundji E, Walker J.
\newblock Discrete choice with social and spatial network interdependencies: an
  empirical example using mixed GEV models with field and panel effects.
\newblock Transportation Research Record: Journal of the Transportation
  Research Board. 2005;1921:70--78.

\bibitem{liben-nowell_2005_geographic}
Liben-Nowell D, Novak J, Kumar R, Raghavan P, Tomkins A.
\newblock Geographic routing in social networks.
\newblock Proceedings of the National Academy of Sciences of the United States
  of America. 2005;102(11623--11628).

\bibitem{carrasco_2008_collecting}
Carrasco JA, Hogan B, Wellman B, Miller EJ.
\newblock Collecting social network data to study social activity- travel
  behaviour: an egocentric approach.
\newblock Environment and Planning B: Planning and Design. 2008;35:961--980.

\bibitem{carrasco_2008_agency}
Carrasco JA, Hogan B, Wellman B, Miller EJ.
\newblock Agency in social activity and ICT interactions: The role of social
  networks in time and space.
\newblock Tijdschrift voor economische en sociale geografie. 2008;99:562--583.

\bibitem{lambiotte_2008_geographical}
Lambiotte R, Blondel VD, De~Kerchove C, Huens E, Prieur C, Smoreda Z, et~al.
\newblock Geographical dispersal of mobile communication networks.
\newblock Physica A: Statistical Mechanics and its Applications.
  2008;387:5317--5325.

\bibitem{krings_2009_urban}
Krings G, Calabrese F, Ratti C, Blondel VD.
\newblock Urban gravity: a model for inter-city telecommunication flows.
\newblock Journal of Statistical Mechanics: Theory and Experiment.
  2009;2009(07):L07003.

\bibitem{phithakkitnukoon_2012_socio}
{Phithakkitnukoon} S, {Smoreda} Z, {Olivier} P.
\newblock Socio-{{Geography}} of {{Human Mobility}}: {{A Study Using
  Longitudinal Mobile Phone Data}}.
\newblock PLoS ONE. 2012;7(6):e39253.

\bibitem{carrasco_2008_how}
Carrasco J, Miller E, Wellman B.
\newblock How far and with whom do people socialize? Empirical evidence about
  the distance between social network members.
\newblock Transportation Research Record: Journal of the Transportation
  Research Board. 2008;2076:114--122.

\bibitem{vandenberg_2013_path}
van~den Berg P, Arentze T, Timmermans H.
\newblock A path analysis of social networks, telecommunication and social
  activity--travel patterns.
\newblock Transportation Research Part C: Emerging Technologies.
  2013;26:256--268.

\bibitem{de_2013_interdependence}
De~Domenico M, Lima A, Musolesi M.
\newblock Interdependence and predictability of human mobility and social
  interactions.
\newblock Pervasive and Mobile Computing. 2013;9(6):798--807.

\bibitem{paez_2007_social}
P{\'a}ez A, Scott DM.
\newblock Social influence on travel behavior: a simulation example of the
  decision to telecommute.
\newblock Environment and Planning A. 2007;39:647--665.

\bibitem{arentze_2008_social}
Arentze T, Timmermans H.
\newblock Social networks, social interactions, and activity-travel behavior: a
  framework for microsimulation.
\newblock Environment and Planning B: Planning and Design.
  2008;35(6):1012--1027.

\bibitem{carrasco_2009_social}
Carrasco JA, Miller EJ.
\newblock The social dimension in action: a multilevel, personal networks model
  of social activity frequency.
\newblock Transportation Research Part A: Policy and Practice. 2009;43:90--104.

\bibitem{hackney_2011_coupled}
Hackney J, Marchal F.
\newblock An agent model of social network and travel behavior interdependence.
\newblock Transp Res Part A. 2011;45:296--309.

\bibitem{ronald_2012_modeling}
Ronald N, Arentze T, Timmermans H.
\newblock Modeling social interactions between individuals for joint activity
  scheduling.
\newblock Transportation research part B: methodological. 2012;46:276--290.

\bibitem{sharmeen_2014_dynamics}
Sharmeen F, Arentze T, Timmermans H.
\newblock Dynamics of face-to-face social interaction frequency: role of
  accessibility, urbanization, changes in geographical distance and path
  dependence.
\newblock Journal of Transport Geography. 2014;34:211--220.

\bibitem{gonzalez_2006_system}
Gonz\'{a}lez MC, Lind P, Herrmann H.
\newblock System of Mobile Agents to Model Social Networks.
\newblock Physical Review Letters. 2006;96(8):088702.

\bibitem{grabowicz_2014_entangling}
Grabowicz PA, Ramasco JJ, Gon{\c{c}}alves B, Egu{\'\i}luz VM.
\newblock Entangling mobility and interactions in social media.
\newblock PLoS One. 2014;9(3):e92196.

\bibitem{toole_2015_coupling}
{Toole} JL, Herrera-Yag\"ue C, Schneider CM, {Gonz{\'a}lez} MC.
\newblock Coupling social mobility and social ties.
\newblock Journal of The Royal Society Interface. 2015;12:20141128.

\bibitem{stouffer_1940_intervening}
Stouffer SA.
\newblock Intervening Opportunities: A Theory Relating Mobility and Distance.
\newblock American Sociological Review. 1940;5(6):845--867.

\bibitem{heanus_1966_comparative}
Heanue KE, Pyers CE.
\newblock A comparative evaluation of trip distribution procedures.
\newblock Highway Research Record. 1966;114:20--50.

\bibitem{ruiter_1967_toward}
Ruiter ER.
\newblock Toward a better understanding of the intervening opportunities model.
\newblock Transportation Research. 1967;1:47--56.

\bibitem{haynes_1973_intermetropolitan}
Haynes KE, Poston~Jr DL, Schnirring P.
\newblock Intermetropolitan Migration in High and Low Opportunity Areas:
  Indirect Tests of the Distance and Intervening Opportunities Hypotheses.
\newblock Economic Geography. 1973;49(1):68--73.

\bibitem{wilson_1970_urban}
Wilson AG.
\newblock Urban and regional models in geography and planning.
\newblock Wiley, New York; 1970.

\bibitem{fik_1990_spatial}
Fik TJ, Mulligan GF.
\newblock Spatial flows and competing central places: Toward a general theory
  of hierarchical interaction.
\newblock Environment and Planning A. 1990;22:527--549.

\bibitem{akwawua_2001_development}
Akwawua S, Poller JA.
\newblock The development of an intervening opportunities model with spatial
  dominance effects.
\newblock Journal of Geographical Systems. 2001;3:69--86.

\bibitem{simini_2012_universal}
Simini F, Gonz{\'a}lez MC, Maritan A, Barab{\'a}si AL.
\newblock A universal model for mobility and migration patterns.
\newblock Nature. 2012;484(7392):96--100.

\bibitem{simini_2013_human}
Simini F, Maritan A, N{\'e}da Z.
\newblock Human Mobility in a Continuum Approach.
\newblock PLoS ONE. 2013;8(3):e60069.

\bibitem{yang_2014_limits}
Yang Y, Herrera C, Eagle N, Gonz{\'a}lez MC.
\newblock Limits of Predictability in Commuting Flows in the Absence of Data
  for Calibration.
\newblock Scientific Reports. 2014;4(5662).

\bibitem{carra_2016_modeling}
Carra G, Mulalic I, Fosgerau M, Barthelemy M.
\newblock Modeling the relation between income and commuting distance.
\newblock Journal of the Royal Society Interface. 2016;13:20160306.

\bibitem{zipf_1946_p1}
Zipf GK.
\newblock {The P1 P2/D Hypothesis: On the Intercity Movement of Persons}.
\newblock American Sociological Review. 1946;11(6):677--686.

\bibitem{barthelemy_2011_spatial}
Barthelemy M.
\newblock Spatial Networks.
\newblock Physics Reports. 2011;499:1--101.

\bibitem{erlander_1990_gravity}
Erlander S, Stewart NF.
\newblock The Gravity model in transportation analysis: theory and extensions.
\newblock Topics in transportation. Utrecht, The Netherlands: VSP; 1990.

\bibitem{ortuzar_2011_modeling}
de~Dios~Ort{\'u}zar J, Willumsen L.
\newblock Modeling Transport.
\newblock New York: John Wiley and Sons Ltd; 2011.

\bibitem{karemera_2000_gravity}
Karemera D, Oguledo VI, Davis B.
\newblock A gravity model analysis of international migration to North America.
\newblock Applied Economics. 2000;32(13):1745--1755.

\bibitem{patuelli_2007_network}
Patuelli R, Reggiani A, Gorman SP, Nijkamp P, Bade FJ.
\newblock Network analysis of commuting flows: A comparative static approach to
  German data.
\newblock Networks and Spatial Economics. 2007;7:315--331.

\bibitem{wilson_1970_entropy}
Wilson AG.
\newblock Entropy in urban and regional modelling.
\newblock London: Pion; 1970.

\bibitem{lenormand_2015_systematic}
Lenormand M, Bassolas A, Ramasco JJ.
\newblock Systematic comparison of trip distribution laws and models.
\newblock Journal of Transport Geography. 2016;51:158--169.

\bibitem{wilson_family_1971}
Wilson AG.
\newblock A {Family} of {Spatial} {Interaction} {Models}, and {Associated}
  {Developments}.
\newblock Environ Plan A. 1971;3(1):1--32.
\newblock Available from: \url{https://doi.org/10.1068/a030001}.

\bibitem{hilton_predictive_2020}
Hilton B, Sood AP, Evans TS.
\newblock Predictive limitations of spatial interaction models: a
  non-{Gaussian} analysis.
\newblock Sci Rep. 2020;10(1):17474.
\newblock Available from:
  \url{https://www.nature.com/articles/s41598-020-74601-z}.

\bibitem{masucci_gravity_2013}
Masucci AP, Serras J, Johansson A, Batty M.
\newblock Gravity versus radiation models: {On} the importance of scale and
  heterogeneity in commuting flows.
\newblock Phys Rev E. 2013;88(2):022812.
\newblock Available from:
  \url{https://link.aps.org/doi/10.1103/PhysRevE.88.022812}.

\bibitem{piovani_measuring_2018}
Piovani D, Arcaute E, Uchoa G, Wilson A, Batty M.
\newblock Measuring accessibility using gravity and radiation models.
\newblock Royal Society Open Science. 2018;5(9):171668.
\newblock Available from:
  \url{https://royalsocietypublishing.org/doi/full/10.1098/rsos.171668}.

\bibitem{mazzoli_2019_field}
Mazzoli M, Molas A, Bassolas A, Lenormand M, Colet P, Ramasco JJ.
\newblock Field theory for recurrent mobility.
\newblock Nature Communications. 2019;10:3895.

\bibitem{schlapfer_2021_trip}
Schlapfer M, Dong L, O'Keeffe K, Santi P, Szell M, Salat H, et~al.
\newblock Trip frequency is key ingredient in new law of human travel.
\newblock Nature. 2021;593:522.

\end{thebibliography}

\end{document}